# Tuning Magnetism in Layered Magnet VI$_3$: A Theoretical Study

Ming An,[†] Yang Zhang,[†] Jun Chen,[†] Hui-Min Zhang,[†] Yunjun Guo,[†,‡] and Shuai Dong*,[†]

[†]School of Physics, Southeast University, Nanjing 211189, China

[‡]School of Physics and Electronic Engineering, Jiangsu Second Normal University, Nanjing 210013, China

**Abstract:** When combined with transition metals with partially filled *d*-orbitals, magnetism can be incorporated in two-dimensional materials, which greatly expands the scope for fundamental researches and potential applications of these materials. Here, a first-principles study of a new two-dimensional ferromagnet VI$_3$ has been carried out. The structural symmetry, magnetic and electronic properties of VI$_3$ in its bulk and single layer forms have been confirmed and predicted, respectively. Its ferromagnetic Curie temperature is predicted to be reduced by half in its monolayer form. In addition, the cation substitution in its monolayer have also been studied, which can significantly tune the magnetism.

## ■ INTRODUCTION

Inspired by the exotic properties of graphene, great enthusiasms have been invested in the research field of two-dimensional (2D) materials, which has gained remarkable progress in the past decade. Numerous layered materials such as III-V compounds, transition metal dichalcogenides (TMDs), and many other 2D compounds with a vast range of novel properties have been fabricated and studied intensively. The common feature of these layered materials is the strong in-plane covalent bonds and the relatively weak out-of-plane van der Waals (vdW) interactions, and thus these layers can be easily exfoliated without damaging the extracted layer. Along with the reduction of material dimension, peculiar nature emerges, making them the ideal platforms for exploring interesting fundamental physics and designing various applications such as catalytic, optoelectronic, and spintronic devices.

For the potential spintronic applications, atomically thin layers with long-range magnetic orders are extremely desired. However, intrinsic magnetic order is quite rare in aforementioned III-V compounds and TMDs. Shortly after the recent discovery of 2D ferromagnetism in $CrI_3$ and $Cr_2Ge_2Te_6$ monolayer/few-layer,[1-3] a large number 2D ferromagnets have been reported or predicted,[4-9] opening a new branch in the plentiful field of 2D materials.

Similar to $CrI_3$, vanadium trihalides, i.e. $VX_3$ ($X$=Cl, Br, I), belong to the big family of transition metal trihalides (TMTs) with layered structure, which have been rarely noticed or mentioned until recently.[10] Son *et al.* and Tian *et al*. have successively synthesized and measured the bulk $VI_3$, and both confirmed it as a Mott insulator with intrinsic ferromagnetism.[11,12] However, the detailed crystal structure remains controversial, and the electrical properties are also inconsistent with former theoretical report.[10] In particular, most layered trihalides adopt either the rhombohedral $BiI_3$-type structure with space group $R\bar{3}$,[13-15] or the monoclinic $AlCl_3$-type structure with $C2/m$ symmetry.[8,16,17] However $VI_3$ exhibits $C2/c$ symmetry according to Son's refined XRD data at low temperature, as compared in Fig. 1. The structural difference between these symmetries are mainly the interlayer stacking pattern, and also the uniqueness of in-plane bond lengths. The interlayer stacking mode can significantly change the inter-layer magnetic coupling, as studied in $CrI_3$ bilayer.[18] In addition, although in most of these trihalides the in-plane spacing diversity of metal ions is quite weak, its effect on the magnetic properties may be significant. For example, the tiny dimerization of Ti in $TiCl_3$ can suppress the magnetism and the magnetic ground state is quite subtle against the tiny structural distortions in $RuCl_3$.[19,20] Therefore, theoretical studies are urgently needed, for both bulk and monolayer $VI_3$.

In the present work, a systematic study of the crystal structure, electronic and magnetic properties of the 2D $VI_3$ both in the bulk and monolayer forms has been carried out using first-principles calculations. Our results confirm the experimental reports on ferromagnetism and Mottness of bulk $VI_3$. This intrinsic ferromagnetism can persist down to a single layer but its Curie temperature ($T_C$) will be reduced by half and its magnetocrystalline anisotropy is changed. Moreover, half V-site substitutions have also been tested, which have a significant effect on material properties. Hopefully, this can be of great importance from the point of view of low dimensional magnetism and potential applications, like the CrI$_3$ case.[21]

## ■ COMPUTATIONAL METHOD

The spin-polarized density functional theory (DFT) calculations were performed with the Perdew-Burke-Ernzerhof (PBE) functional in the generalized gradient approximation (GGA) as implemented in the Vienna *ab initio* simulation package (VASP).[22,23] To properly describe the correlated electrons, the GGA+*U* method was used. The on-site Hubbard $U_{\text{eff}}$ was imposed on 3*d* orbital using the Dudarev approach for the structural optimization, electronic structure, and magnetic ground state calculations.[24] The spin-orbit coupling (SOC) effect was also considered during the magnetic calculations, considering the large atom number of iodine. For bulk calculations, the interlayer vdW corrections were considered within the Grimme's approach (DFT-D2).[25] In the study of monolayer structure, a vacuum space of 15 Å was adopted to avoid the interaction between two neighboring slices.

Both the lattice constants and atomic positions were fully relaxed until the force and the energy were converged to 0.01 eV/Å and 10$^{-5}$ eV, respectively. The plane-wave cutoff energy was set to 500 eV. The Monkhorst-Pack *k*-point meshes were chosen as 5×5×2 and 5×3×3 for the rhombohedral and monoclinic structure, respectively. For the monolayer model, only the center point is selected along the $k_z$-direction while the grid densities remain unchanged in other directions.

## ■ RESULTS AND DISCUSSION

### A. Confirmation of bulk's properties

Before the studying of monolayer, it is necessary to check its bulk properties, especially considering the existence of disagreement between previous theoretical work and experimental report.[10-12]

Belonging to the TMT family, VI$_3$ bulk structure was once described in the rhombohedral space group $R\bar{3}$ below 79 K, analogous to its chromium trihalide cousins,[10,12,26] as shown in Fig. 1(a-b). However, based on experimental data, Son *et al.* claimed an usual *C*2/*c* structure [Fig. 1(c-d)] for VI$_3$ bulk below 40 K, rather than $R\bar{3}$.[11]

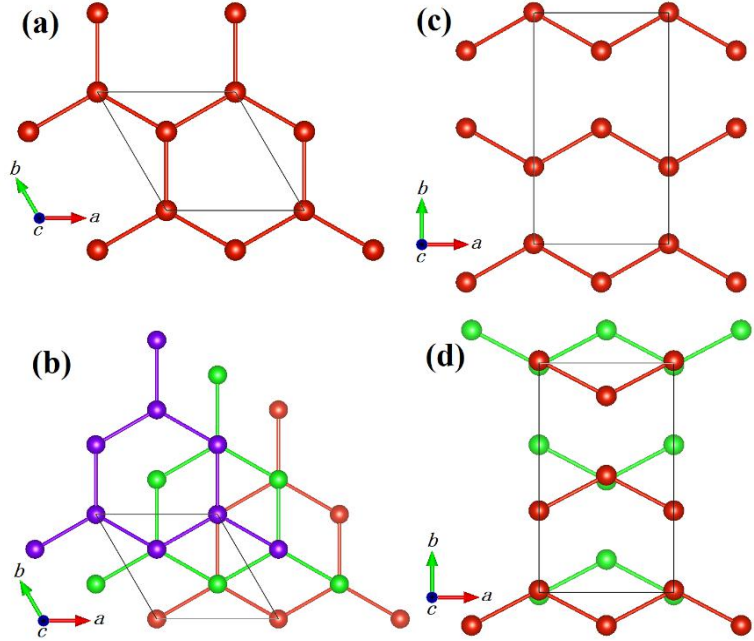

**Figure 1.** Schematic of (a-b) the $R\bar{3}$ and (c-d) *C*2/*c* structures. I atoms are omitted for clarify. (a) and (c) are the top views of single layers of each structure, while (b) and (d) are the perspective view of one unit cell (different colors denote different layers). There are two key differences between the $R\bar{3}$ and *C*2/*c* structure. First, for the $R\bar{3}$ structure, the stacking mode is the *ABC*-type, while it is *AA'* mode for the *C*2/*c*. Second, the *AA'* stacking mode breaks the $C_3$ rotation symmetry, thus the in-plane V-V bond lengths are no longer unique in the *C*2/*c* structure.

To resolve this uncertainty, the crystal structure of bulk VI$_3$ is determined in our calculation. By comparing the total energies of fully optimized *C*2/*c* and $R\bar{3}$ phase, the *C*2/*c* structure is found to be more favorable: 6.8 meV/V lower than that of $R\bar{3}$. In the monoclinic *C*2/*c* phase, the slightly staggered layers separated by vdW gap lead to a slightly larger neighbor I-I interlayer distance, resulting in the reduction of interlayer Coulombic repulsion and the total energy of *C*2/*c* structure. In addition to the different stacking pattern along *c* axis, the in-plane difference between two phases is subtle. In particular, the V ions in the *C*2/*c* phase spontaneously shift along a single direction within the *ab* plane, breaking the $C_3$ rotation symmetry and tending to form zigzag chains perpendicular to the displacement direction, in agreement with the experimental result.[11] The difference of in-plane neighboring V ion spacing is in the order of 0.01 Å (4.003 Å vs 4.010 Å) in the *C2/c* phase, while the in-plane V-V distance is uniform in the $R\bar{3}$ phase (4.015 Å).

In a previous theoretical report,[10] VI$_3$ was predicted to be a Dirac half-metal with a gap larger than 4 eV in one spin channel and a Dirac cone in the other. While according to recent optical measurement, a direct band gap of 0.67 eV was observed,[11,12] suggesting that VI$_3$ is most likely a correlated Mott insulator.

Here the density of states (DOS) for bulk VI$_3$ in the ferromagnetic (FM) state has been calculated, as shown in Fig. 2(a). The $U_{\text{eff}}$ dependent band gap is summarized in Fig. 2(b). Our calculation confirms that the *C2/c* and $R\bar{3}$ phases have very similar electronic structures, thus the latter of which is not shown here. The valence band maximum (VBM) are mainly contributed by I's 5*p* and V's 3*d* orbitals, while the conduction band minimum (CBM) are dominant by V's 3*d* orbitals. Obviously, when the Hubbard $U_{\text{eff}}$ is around 3 eV, the electronic structure of VI$_3$ bulk can be well captured with a moderate bandgap of 0.7 eV, which is in good agreement with the experimental value. Besides, this particular $U_{\text{eff}}$ value is close to the parameter evaluated by linear response theory.[27] Therefore, $U_{\text{eff}}$=3 eV will be used in the following calculations, if not otherwise explicitly noted.

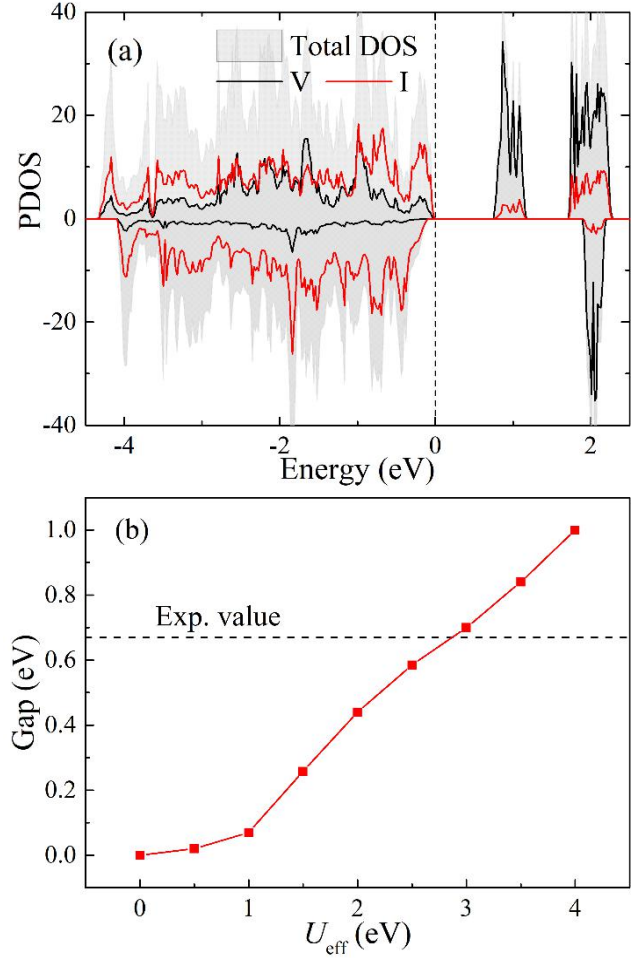

**Figure 2**. (a) DOS and atomic-projected DOS (PDOS) of VI$_3$ bulk with $C2/c$ symmetry with $U_{eff}$ = 3 eV. (b) Diagram of the $U_{eff}$ dependent band gap.

When cooled down to 50 K, a long-range ferromagnetism was observed in VI$_3$ bulk by several experimental groups.[11,12,26] To investigate the magnetic property of bulk VI$_3$ with $C2/c$ symmetry at low temperature, the spin-polarized DFT calculations has been carried out. The magnetic ground state is determined by comparing the relative energies among four most typical magnetic configurations in honeycomb lattice, i.e. the FM, the checkerboard Néel antiferromagnetic (nAFM), the stripy antiferromagnetic (sAFM), and the zigzag antiferromagnetic (zAFM) states. More detailed description about these four distinct configurations can be found in Ref. 10. The SOC effect is included during the magnetic calculations. Our calculation confirms that the FM state is the most stable magnetic configuration, in consistent with the experimental reports.[11,12] The local magnetic moment of V is 1.98 μ$_B$, as expected for the high-spin state of V$^{3+}$ cation with the 3$d^2$ ($S$=1) configuration, and the orbital moment is zero. The first excited state is the sAFM state which is 18.3 meV/V higher than the FM ground state, while the nAFM and zAFM are almost energetically degenerate and 12.8 meV/V higher than that of sAFM. The $R\bar{3}$ structure has also been checked, which shows not effect on the FM ground state. The only difference is the first excited state, which changes from sAFM for C2/c phase to nAFM for $R\bar{3}$. Moreover, the robust stability of FM ground state is further confirmed by using different pseudopotential, e.g., PBEsol functional.

To further understand the magnetism, a simplified Heisenberg model on a 2D honeycomb lattice is proposed as:

$$H = J\sum_{<i,j>} S_i \cdot S_j + K\sum_{i}(S_i^z)^2 , \qquad (1)$$

where $J$ is the exchange interaction between nearest neighbor V spins ($S$) and $K$ is the coefficient of magnetocrystalline energy. As a preliminary model, the inter-layer coupling, intra-layer difference of $J$, the couplings beyond nearest neighbors are all neglected considering the long distance between ions and the vdW type coupling. Therefore, the two-dimensional Heisenberg model is adopted to simulate the magnetism for bulk. Based on the energy difference between FM and n-AFM states, the estimated value of $J$ is -10.7 meV. Meanwhile, the coefficient $K$ is also estimated as -0.97 meV, implying an easy-axis along the $c$-axis, in agreement with the experimental observations.[11,12] Then the FM $T_C$ is estimated as ~60 K, accroding to our Monte Carlo simulation, which is close to the reported experimental value 50 K.[12]

### B. Properties of monolayer

After the successful confirmation of bulk's properties, it is safe to explore the properties of monolayer. To evaluate the feasibility of exfoliation from VI$_3$ bulk, the binding energy $E_B$ was calculated by gradually increasing the interlayer spacing $d$. As shown in Fig. 3(a), $E_B$ increases quickly as separation $d$ increases and then converges to the saturation value 0.25 J/m$^2$ when $d$ is larger than 8 Å. Our result is between the former theoretical values: 0.18 and 0.29 J/m$^2$,[10,12] indicating the rationality of our calculation. This estimated value is also comparable with those of Cr$X_3$ materials [21,28] and in the same order of magnitude comparing with other previous reported 2D materials.[29] Thus, the weak interlayer interaction in VI$_3$ leads its easy exfoliation from its bulk phase.

Then the question arises: can FM ordering persist with dimensional reduction down to monolayer, just like the CrI$_3$ case?[3] To find out the answer, the isolated monolayer is calculated. First, the in-plane lattice constants have been obtained through structural optimization. Noting that for a single layer, the stacking issue as discussed in Fig. 1, does not exist any more. Then the $C_3$ rotation symmetry is restored. The optimized lattice constant is 7.16 Å, which is ~3% larger than the bulk value. Meanwhile the thickness of monolayer, i.e. the distance between upper and lower atomic planes of iodine, is reduced by 1%. In other words, the VI$_3$ monolayer will spontaneously stretch out when exfoliated from its bulk phase, which is reasonable considering the absence of interlayer vdW attraction. The electronic structure of monolayer VI$_3$ has been calculated and shown in Fig. 3(b). The DOS of monolayer VI$_3$ is similar to their bulk counterparts except for the fact that the band gap is a little larger (~ 0.89 eV), which is due to the in-plane extension. Such expansion increases the V-I bond lengths, leads to weaker *p-d* hybridization and thus stronger electron localization. Similar phenomenon is also found in other 2D transition metal compounds.[30]

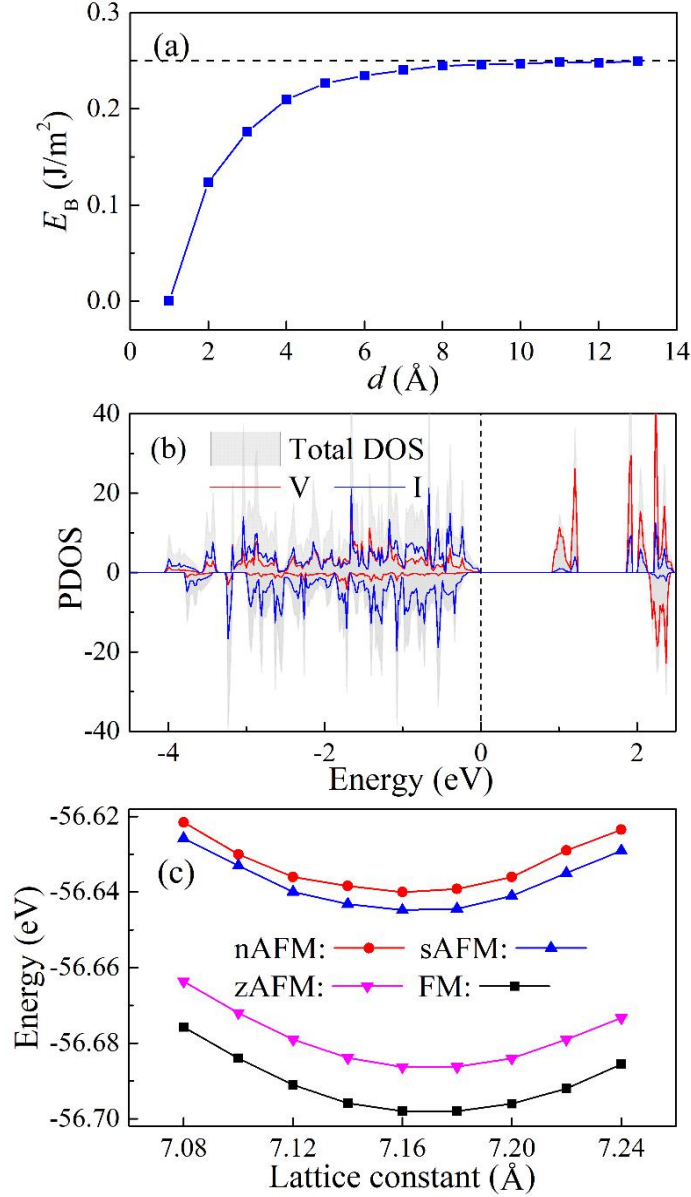

**Figure 3**. (a) Binding energy $E_B$ as a function of the interlayer separation $d$. (b) PDOS of VI$_3$ single layer with $C2/m$ symmetry. (c) Energy comparison of different magnetic configurations. For the sake of simplicity, the results are calculated with spin moments pointing out-of-plane, since the relative energies among different configurations are barely affected by the specific direction of magnetic moments. The $U_{eff}$ = 3 eV is adopted for all above calculations.

Such in-plane expansion can significantly tune the magnetism since the exchange interaction seriously depends on bond lengths as well as bond angles. Our spin-polarized calculation of VI$_3$ monolayer confirms that the FM state remains the lowest energy one, with a local moment of 2.01 $\mu_B$ per V. Such a FM ordering remains unchanged upon the application of biaxial stress, as shown in Fig. 3(c). However, the exchange interaction $J$ in monolayer is significantly reduced, which is only ~-5.0 meV. Then the FM $T_C$ simulated by Monte Carlo simulation is only ~27 K, reduced by half comparing with the bulk value. Additionally, the magnetocrystalline coefficient is also changed a lot, which is 0.77 meV/V, implying an easy plane in monolayer in opposite to the easy-axis in bulk.

Further experiments on VI$_3$ monolayer are encouraged to verify our prediction regarding the reduced FM $T_C$ as well as the change of magnetocrystalline anisotropy.

## C. Effect of substitution

In the following, the effect of cation substitution on V site will be investigated to tune the electronic and magnetic properties. As the first step, here we only consider the simplest substitution, i.e. by replacing half of V by Ti, Mn, and Fe orderly, as shown in Fig. 4(a). The effects of on-site Hubbard interaction are considered. The effective parameter $U_{eff}$ are set to 0.8 eV, 2.5 eV, and 4.0 eV for Ti, Mn, and Fe elements, respectively, according to previous literature.[31,32] To examine the chemical stability of VTiI$_6$, VMnI$_6$, and VFeI$_6$ monolayers, their formation energies are first estimated, which can be expressed by the formula $E_f = E(VMI_6) - E(VI_3) - E(MI_3)$, where $E(VMI_6)$, $E(VI_3)$, and $E(MI_3)$ are the energy of V$M$I$_6$, VI$_3$, and $M$I$_3$ ($M$ = Ti, Mn, and Fe), respectively. The calculated $E_f$ for VTiI$_6$, VMnI$_6$, and VFeI$_6$ monolayers are -0.89 eV/f.u., -1.61 eV/f.u., and -0.41 eV/f.u., respectively. The negative values indicate that their chemical stabilities are probably allowed. Additionally, the phonon dispersion calculations and molecular dynamical simulations (Fig. S1 in Supporting Information) have been carried out, which demonstrate their dynamical and thermal stabilities.

The lattice constants obtained by optimization are 7.20 Å, 7.15 Å, and 7.10 Å for VTiI$_6$, VMnI$_6$, and VFeI$_6$, respectively. The decreasing lattice constant with the increasing atomic number of substituted element is reasonable, considering the shrinking tendency of ion size. The strain-dependent magnetic properties of these substituted materials are also calculated, as shown in Fig. 4(b-d). The VTiI$_6$ monolayer remains FM and none AFM state exist. The reason is that there is a complete charge transfer from Ti to V. Then the valence of Ti ion becomes +4, without 3$d$ electrons (i.e. nonmagnetic). Meanwhile, the valence of V ion becomes +2, with the 3$d^3$ configuration. For VMnI$_6$ single layer, the total energy of FM configuration is 7.9 meV/f.u. lower than that of nAFM phase. This energy advantage of FM ground state is slightly larger than that of VI$_3$ monolayer, and will be suppressed by both tensile and compressive strains. In contrast, the nAFM order becomes the ground state in VFeI$_6$ with a sizable energy advantage which is robust against strain. The V's and Fe's spins are antiferromagnetically coupled with local moments of -2.0 μ$_B$/V and 3.8 μ$_B$/Fe, respectively. Thus, it is a ferrimagnetic material and its Néel temperature is ~41 K according to our Monte Carlo simulation. Due to the partial covalent bonds, the iodine ions are also partially magnetic polarized, with a magnetic moment of 0.2 μ$_B$/I parallel to iron's. Consequently, the net magnetization for VFeI$_6$ monolayer is 3.0 μ$_B$/f.u..

The band structure and DOS of these substituted materials are shown in Fig. 5. The undoped VI$_3$ monolayer is also given for comparison. For VTiI$_6$ monolayer, the 3$d$ unoccupied states from Ti ions dominate the CBM, leading to a band gap slightly narrower than that of VI$_3$. For VMnI$_6$ monolayer, the hybridized Mn's 3$d$ and I's 5$p$ states mainly contribute to the VBM and lead to the metallic behavior in spin-down channel. While a band gap of 2.3 eV opens in the spin-up channel presenting half-metallic property, which has great advantages for spintronic applications. In the case of VFeI$_6$, the insulating characteristic is preserved and the band gap is almost identical to that of VI$_3$ monolayer, since the strong hybridizations between Fe's 3$d$ and I's 5$p$ orbitals take place far below the Fermi level.

## ■ CONCLUSION

In summary, we have investigated the structure, electronic and magnetic properties of VI$_3$ in its bulk and monolayer forms. Our results confirmed that VI$_3$ is a Mott-insulator and an intrinsic ferromagnetic material with saturated magnetization 2.0 μ$_B$/V. Its structure symmetry is $C2/c$ instead of the more popular $R\bar{3}$. VI$_3$ monolayers can be easily exfoliated from the bulk. The ferromagnetism can be maintained down to monolayer but its Curie temperature will be reduced by half and the magnetocrystalline anisotropy will be changed. Furthermore, our calculations also suggest that the magnetic and electronic properties can be modulated by V-site substitution. Our study provides a comprehensive understanding of the nature of VI$_3$ and will hopefully stimulate further experimental effort in VI$_3$ and its derivations.

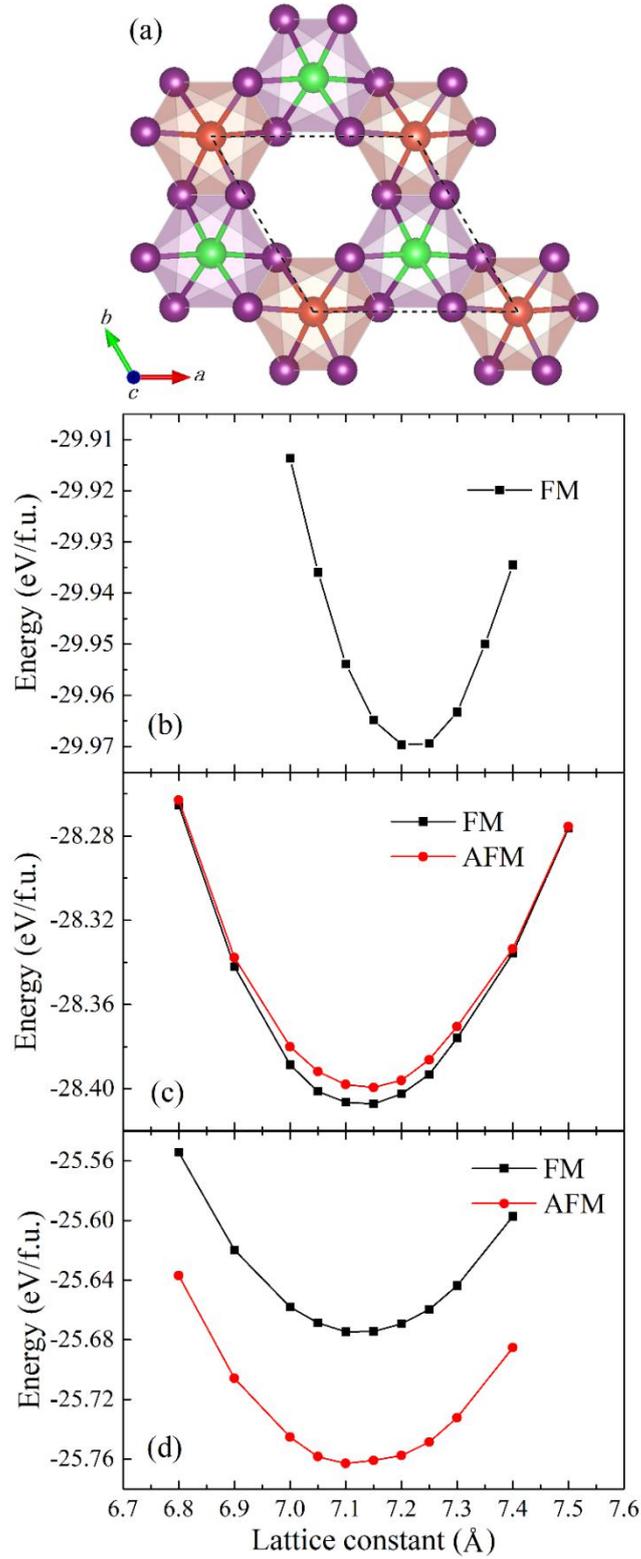

**Figure 4**. (a) Schematic of V$M$I$_6$ ($M$ =Ti, Mn, and Fe). The triangular lattice is represented by dashed lines, each containing one formula unit (f.u.). (b-d) Energy comparison between FM and nAFM states in (a) VTiI$_6$, (b) VMnI$_6$, and (c) VFeI$_6$ monolayers, respectively.

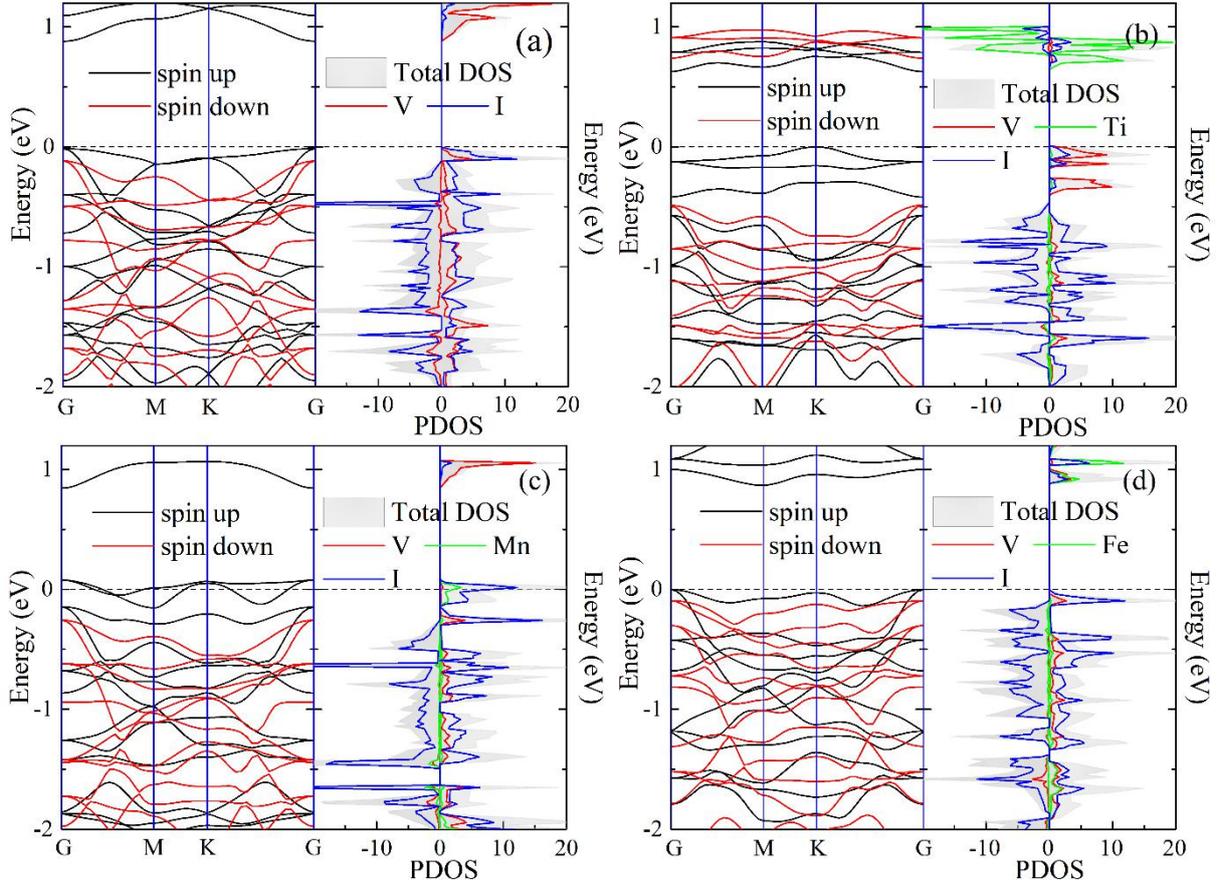

**Figure 5**. The electronic structures (bands, DOS, PDOS) of (a) $VI_3$, (b) $VTiI_6$, (c) $VMnI_6$, and (d) $VFeI_6$ monolayers. The effective Hubbard $U_{eff}$ adopted for V, Ti, Mn, and Fe are 3.0, 0.8, 2.5, and 4.0 eV, respectively.

■ **AUTHOR INFORMATION**

**Corresponding Author**

*E-mail: sdong@seu.edu.cn

**Notes**

The authors declare no competing financial interest.

■ **ACKNOWLEDGMENTS**

This work was supported by National Natural Science Foundation of China (Grant Nos.13